\documentclass[floatfix,pra,twocolumn,aps,superscriptaddress,showpacs]{revtex4-1}
\usepackage{hyperref}
\hypersetup{
	colorlinks=true,
	citecolor=blue,
	linkcolor=blue,
	urlcolor=blue}
\usepackage{graphicx}
\usepackage[normalem]{ulem}
\usepackage{amsmath}
\usepackage{amsfonts}
\usepackage{amssymb}
\usepackage{epsfig}
\usepackage{subfigure}
\usepackage{mathtools}
\usepackage{braket}
\usepackage[usenames,dvipsnames]{color}
\usepackage{setspace}
\usepackage{bm}
\usepackage{times}
\usepackage{ulem}
\usepackage{dsfont}
\usepackage{xcolor}

\begin{document}
\title{Enhanced abrupt autofocusing and depth-of-focus of circular Airy derivative beams \\via controlled asymmetry}
\author{Anita Kumari} 
\affiliation{Department of Physics, Indian Institute of Technology Ropar, Rupnagar 140001, Punjab, India}
\author{Vishwa Pal}\email{vishwa.pal@iitrpr.ac.in}
\affiliation{Department of Physics, Indian Institute of Technology Ropar, Rupnagar 140001, Punjab, India}

\begin{abstract}
We demonstrate simultaneous enhancement of abrupt autofocusing and depth-of-focus (DOF) in circular Airy derivative beams through controlled asymmetry. The controlled asymmetry enables multiple discrete autofocusing events to merge into a single elongated focal region, thereby enhancing both the autofocusing ability and the DOF. The DOF is extended by up to $400\thinspace\%$ with a $32\thinspace\%$ improvement in autofocusing ability and a $\sim32\,\mu$m focal spot size, whereas the autofocusing ability can be further enhanced by up to $193\thinspace\%$ with a $\sim56\,\mu$m focal spot size, while maintaining nearly the same DOF. Experimental results show good agreement with numerical simulations. The resulting asymmetric circular Airy derivative beams (ACADBs), combining strong abrupt autofocusing with extreme DOF, offer promising applications in optical trapping, optical communications, microfabrication, and biomedical imaging.
\end{abstract}
\maketitle
\section{Introduction}
Optical beams with strong focusing and extended depth-of-focus (EDOF) have attracted considerable interests for a wide range of applications, including optical trapping \cite{zhang2011trapping}, light-sheet microscopy \cite{kafian2020light}, optical coherence tomography \cite{lorenser2012ultrathin}, biomedical imaging \cite{planchon2011rapid}, filamentation \cite{fu2024extending}, and free-space optical communication \cite{killinger2002free}. In this context, non-diffracting beams$-$such as plane waves, Bessel beams \cite{khonina2020bessel}, Airy beams \cite{efremidis2019airy}, needle-shaped beams \cite{zhao2022flexible}, and pin-like beams \cite{hu2022experimental}, have been extensively explored. However, their inherently large transverse extent and pronounced sidelobes often constrain their applicability. 

Achieving a tightly confined optical field at a desired location is traditionally realized using conventional focusing elements such as lenses, mirrors, or diffractive optical components. 
However, such approaches often suffer from several limitations, including sensitivity to precise mechanical alignment and rapid diffraction-induced spreading of the beam beyond the focal region. Moreover, the gradual focusing process may lead to unwanted interaction between the beam and the surrounding medium before reaching the focal point, thereby reducing the effective intensity delivered to the target region. 

To overcome these challenges, considerable attention has been directed toward abruptly autofocusing beams \cite{Papazoglou:11}. Unlike conventional focusing methods, these beams naturally redistributes their energy during propagation, forming a sharp focal spot at a predetermined axial position without the need for external focusing elements. This unique property results in low pre-focal intensity followed by a sudden and pronounced concentration of energy, making them highly suitable for controlled material interactions and localized nonlinear excitation. Over the years, various types of abrupt autofocusing beams have been explored, including the Pearcey beam \cite{Ring:12}, circular Pearcey beam \cite{Chen:18}, circular Pearcey-Gaussian beam \cite{Chen2:18}, circular Airy beam \cite{Papazoglou:11,lu2019abruptly}, modified circular Airy beam \cite{jiang2015propagation}, circular Airy derivative beam (CADB) \cite{zang2022abruptly} and ring Airyprime beams array \cite{Zang:22}. Owing to their strong abrupt autofocusing capability and robust propagation characteristics, CADBs have attracted considerable research interest, leading to a wide range of investigations. These studies include chirp-controlled autofocusing ability and focal length \cite{zang2023simultaneously,he2023key}, tunable autofocusing via Fourier-space modulation \cite{zheng2024adjustable}, as well as autofocusing and self-healing of partially blocked CADBs in both free space and complex media \cite{kumari2024autofocusing,he2024propagation,kumari2025abrupt,kumari2026abrupt}.

Tunable focal length has been demonstrated in ring Airyprime beams \cite{he2023key,zang2023simultaneously}, however, the axial intensity profile is dominated by a sharp primary focus followed by secondary focal events \cite{kumari2025abrupt}. This leads to alternating bright and dark regions, rather than a continuous high-intensity. Such behavior limits their utility in applications that require a sustained high-intensity over an extended axial range with invariant focal spot size. To address this limitation and achieve an EDOF while simultaneously enhancing the autofocusing performance, we introduce a controlled asymmetry into CADBs by imposing complex transverse shifts in Cartesian coordinates. We refer to these beams as asymmetric circular Airy derivative beams (ACADBs). The introduction of complex shifts deliberately breaks the radial symmetry in a controlled manner, redistributing the discrete subsidiary focal events into a more continuous and extended high-intensity region, while also improving the overall autofocusing characteristics.

We present both experimental and numerical results demonstrating ACADBs with EDOF for different degrees of introduced asymmetry, along with enhanced autofocusing ability, reduced focal spot size, and improved on-axis intensity uniformity.

\section{Theoretical description} \label{theo_desc}
\noindent
The normalized electric field of ACADBs in the initial plane ($z=0$) is described as
\begin{equation}
E_{\mathrm{c}}(r,\phi,0) = E_{\mathrm{a1}} + E_{\mathrm{a2}},
\label{Eq:1}
\end{equation}
where the individual fields $E_{\mathrm{a1}}$ and $E_{\mathrm{a2}}$ are expressed as
\begin{equation}
E_{\mathrm{aj}} = Ai'\!\left(\frac{r_0 - r_j}{w_0}\right) \exp\!\left[a\frac{r_0 - r_j}{w_0}\right].
\label{Eq:2}
\end{equation}
Here, $r = \sqrt{x^2 + y^2}$ and $\phi = \tan^{-1}(y/x)$ denote the radial and azimuthal coordinates, respectively. The shifted radial coordinate is defined as
\[
r_j = \sqrt{(x - x_j)^2 + (y - y_j)^2}, \quad j = 1,2.
\]
where $x_j$ and $y_j$ represent the complex transverse shifts along the $x$~- and $y$~-~directions, respectively; $a$ denotes the exponential truncation factor, $w_0$ is the scaling parameter, $r_0$ is related to the radius of the ACADBs at $z = 0$, and $Ai'$ represents the first-order derivative of the Airy function with respect to its argument.

\begin{figure}[htbp]
\centering
\includegraphics[height = 3.3cm, keepaspectratio = true]{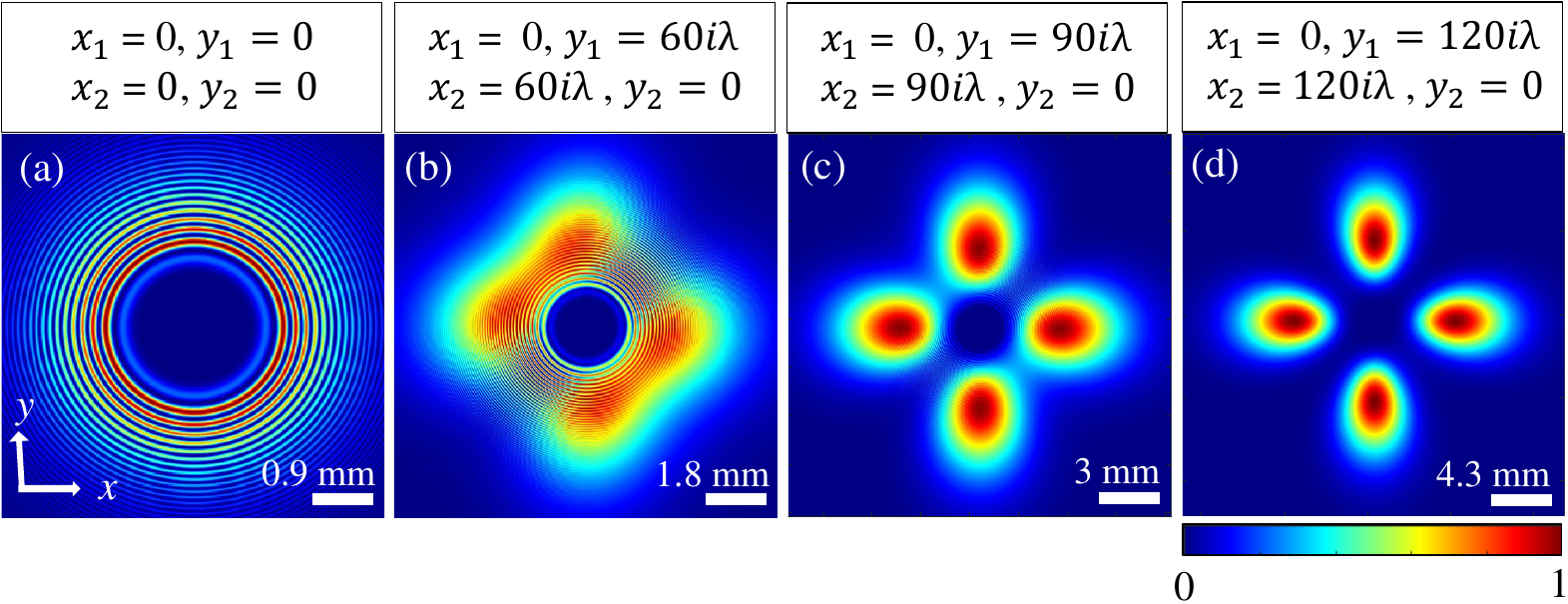}
\caption{Simulated intensity distributions of ACADBs at $z=0$ for different asymmetry cases: (a) A1, (b) A2, (c) A3, and (d) A4, corresponding to progressively increasing complex transverse shifts along orthogonal transverse directions.}
\label{fig:1}
\end{figure}

Figure\,\ref{fig:1} shows the intensity distributions of ACADBs at $z=0$ for different complex transverse shifts. Figure\,\ref{fig:1}(a) represents the symmetric case  A1 ($x_1 = x_2 = y_1 = y_2 = 0$), while Figs.\,\ref{fig:1}(b)~–~\ref{fig:1}(d) show progressively increasing asymmetry: A2 ($x_{1}=0$, $y_1 = 60 i\lambda$, $x_2 = 60 i\lambda$, $y_{2}=0$), A3 ($x_{1}=0$, $y_1 = 90 i\lambda$, $x_2 = 90 i\lambda$, $y_{2}=0$), and A4 ($x_{1}=0$, $y_1 = 120 i\lambda$, $x_2 = 120 i\lambda$, $y_{2}=0$), with the remaining shifts set to zero. As the magnitude of the complex shifts increases, the transverse intensity evolves into a more pronounced petal-like structure.

The propagation of ACADBs along $z$-axis in free space can be calculated using the Fresnel diffraction integral \cite{collins1970}, and solved it numerically using a fast Fourier transform method. Based on experimental convenience, we select the beam parameters as $a = 0.1$, $r_0 = 1\,\mathrm{mm}$, $w_{0} = 0.1\,\mathrm{mm}$, and $\lambda = 1064\,\mathrm{nm}$.

\section{Experimental setup}
\begin{figure}[htbp]
\centering
\includegraphics[height = 4.5cm, keepaspectratio = true]{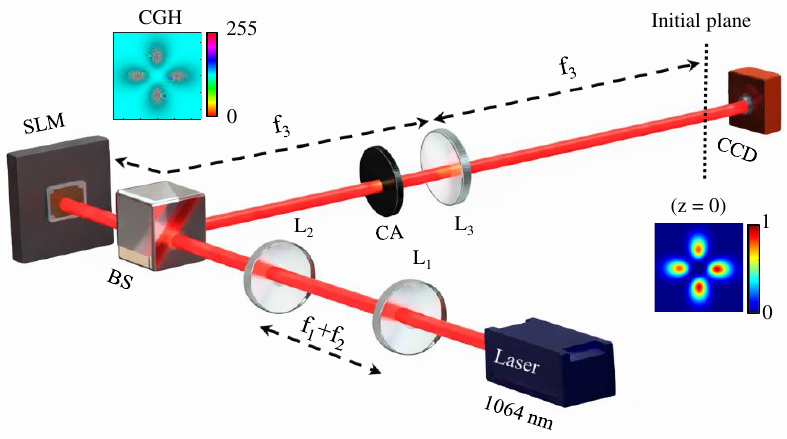}
\caption{Experimental schematic for the generation and free-space propagation of ACADBs. $\mathrm{L}_1$, $\mathrm{L}_2$, and $\mathrm{L}_3$: plano-convex lenses with focal lengths $\mathrm{f}_1$, $\mathrm{f}_2$, and $\mathrm{f}_3$, respectively; BS: Beam splitter; SLM: Spatial light modulator; CGH: Computer generated hologram; CA: Circular aperture; CCD: Camera.}
\label{fig:2}
\end{figure}

The experimental arrangement for the generation and propagation of ACADBs is shown in Fig.\,\ref{fig:2}. A linearly polarized Gaussian laser beam at $\lambda = 1064\,\text{nm}$ is first expanded using a pair of plano-convex lenses, $\mathrm{L}_{1}$ ($\mathrm{f}_{1}=3\,\text{cm}$) and $\mathrm{L}_{2}$ ($\mathrm{f}_{2}=30\,\text{cm}$), arranged in a telescopic configuration. The expanded beam is then directed normally onto a spatial light modulator (SLM) with the assistance of a $50{:}50$ beam splitter (BS). The SLM has a resolution of $1920 \times 1080$ pixels, with each pixel having a size of $8\,\mu\text{m}$. A suitable computer-generated hologram (CGH) is encoded onto the SLM, which modulates both the phase and amplitude of the incident beam (for the details of CGH, see Appendix\,\ref{Appendix_A}). The modulated field emerging from the SLM contains several diffraction orders. To isolate the desired first diffraction order, a circular aperture (CA) is placed to filter out the unwanted orders. A plano-convex lens $\mathrm{L}_{3}$ ($\mathrm{f}_{3}=50\,\text{cm}$) performs the Fourier transformation of the modulated field, resulting in the formation of the desired ACADB at the back focal plane of $\mathrm{L}_{3}$, referred to as the initial plane ($z=0$). The subsequent propagation of the generated beam is recorded using a CCD (charge-coupled device) camera for detailed characterization of its abrupt autofocusing behavior and intensity evolution.

\section{Results and discussion}
\begin{figure*}[htbp]
\centering
\includegraphics[height = 8.4cm, keepaspectratio = true]{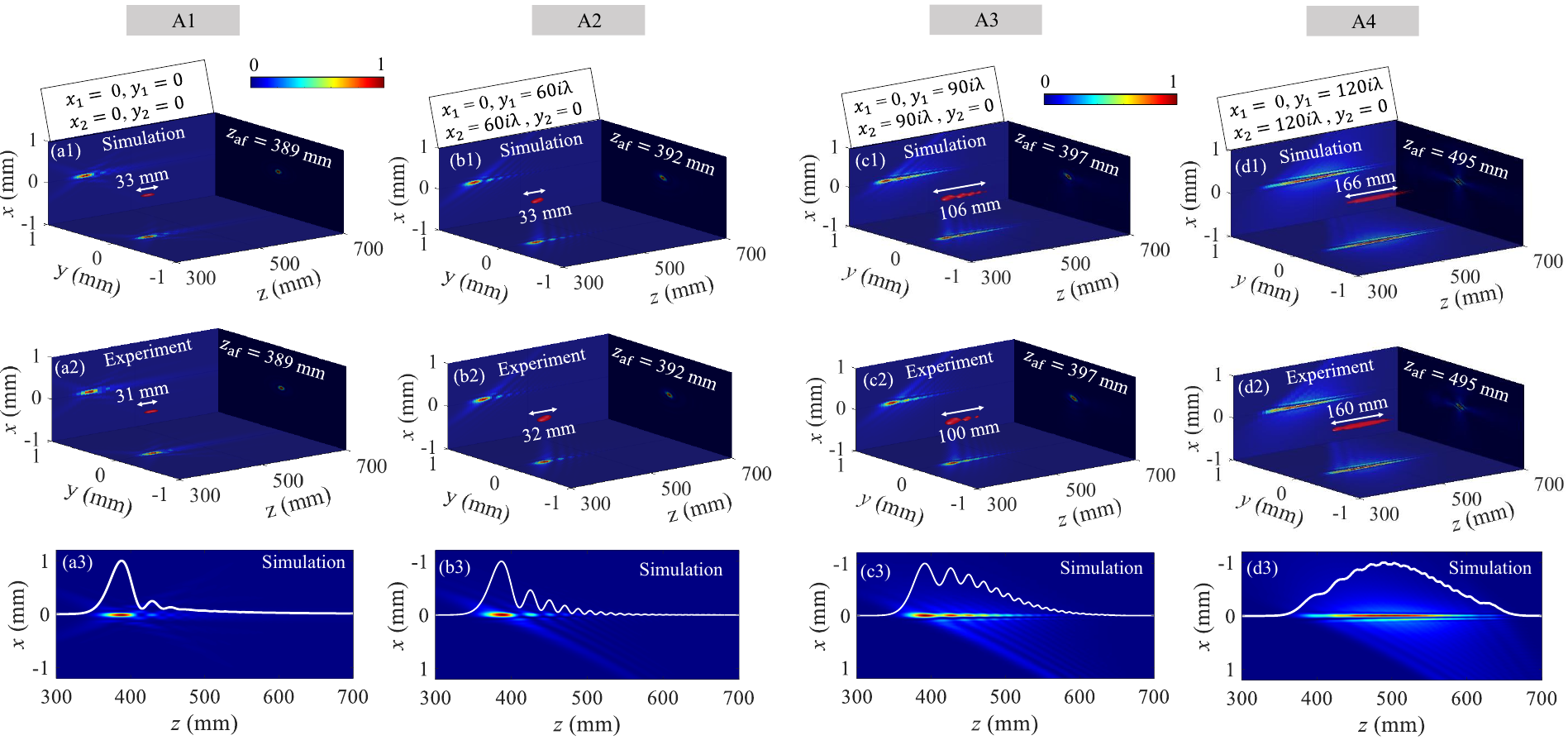}
\caption{Intensity distributions of ACADBs across three orthogonal planes. The red isosurface represents the 3D shape of the focused ACADBs, truncated at $50~\%$ of the peak intensity. (a1, a2), (b1, b2), (c1, c2) and (d1, d2) present simulated and experimental results for A1--A4 cases, respectively. (a3), (b3), (c3) and (d3) show simulated intensity distribution in the $x$--$z$ plane corresponding to the A1--A4 cases, respectively. The white curve represents the on-axis intensity variation.}
\label{fig:3}
\end{figure*}

The abrupt autofocusing ability and EDOF of ACADBs are systematically investigated as the asymmetry in the initial plane increases from A1 to A4 cases. The experimental and numerical results are shown in Figs.\,\ref{fig:3}(a1)--\ref{fig:3}(a3) for A1, Figs.\,\ref{fig:3}(b1)--\ref{fig:3}(b3) for A2, Figs.\,\ref{fig:3}(c1)--\ref{fig:3}(c3) for A3, and Figs.\,\ref{fig:3}(d1)--\ref{fig:3}(d3) for A4. 

For A1 (symmetric case), at $z=0$ (Fig.\,\ref{fig:1}(a)), the CADB exhibits multiple concentric bright and dark rings with a relative phase difference of $\pi$ between adjacent rings. Upon propagation, lateral self-acceleration causes the outer rings to progressively interfere with the inner ones, leading to intensity redistribution toward the beam center. Consequently, at a particular value of $z$ a significant portion of the beam intensity collapses into a tightly localized high-intensity peak (Figs.\,\ref{fig:3}(a1)--\ref{fig:3}(a3)). This position defines the autofocusing distance, $z_{\mathrm{af}}$. The autofocusing mechanism of CADBs can be interpreted through an analogy with a Fresnel zone plate \cite{zang2022abruptly,kumari2024autofocusing}.

As the asymmetry introduced (A2--A4) (Figs.\,\ref{fig:1}(b)--\ref{fig:1}(d)), during propagation the beam intensity again redistributes from the peripheral regions toward the center, forming a distinct autofocusing point at $z=z_{\mathrm{af}}$. Beyond this point, as the asymmetry increases, the intensity of the secondary focal events increases correspondingly. As a result, the beam maintains a focused state over an extended propagation distance, as shown in Figs.\,\ref{fig:3}(b1)--\ref{fig:3}(b3)) for A2, Figs.\,\ref{fig:3}(c1)--\ref{fig:3}(c3)) for A3, and Figs.\,\ref{fig:3}(d1)--\ref{fig:3}(d3)) for A4. Here, the length of the red isosurface, representing the 3D focused ACADBs truncated at
$50\%$ of the peak intensity, increases, with corresponding values of $33\thinspace$mm, $33\thinspace$mm, $106\thinspace$mm, and $166\thinspace$mm for A1–A4, respectively. These results clearly demonstrate the enhanced axial confinement of the focused ACADBs with increasing asymmetry. 

\begin{figure}[htbp]
\centering
\includegraphics[height = 6cm, keepaspectratio = true]{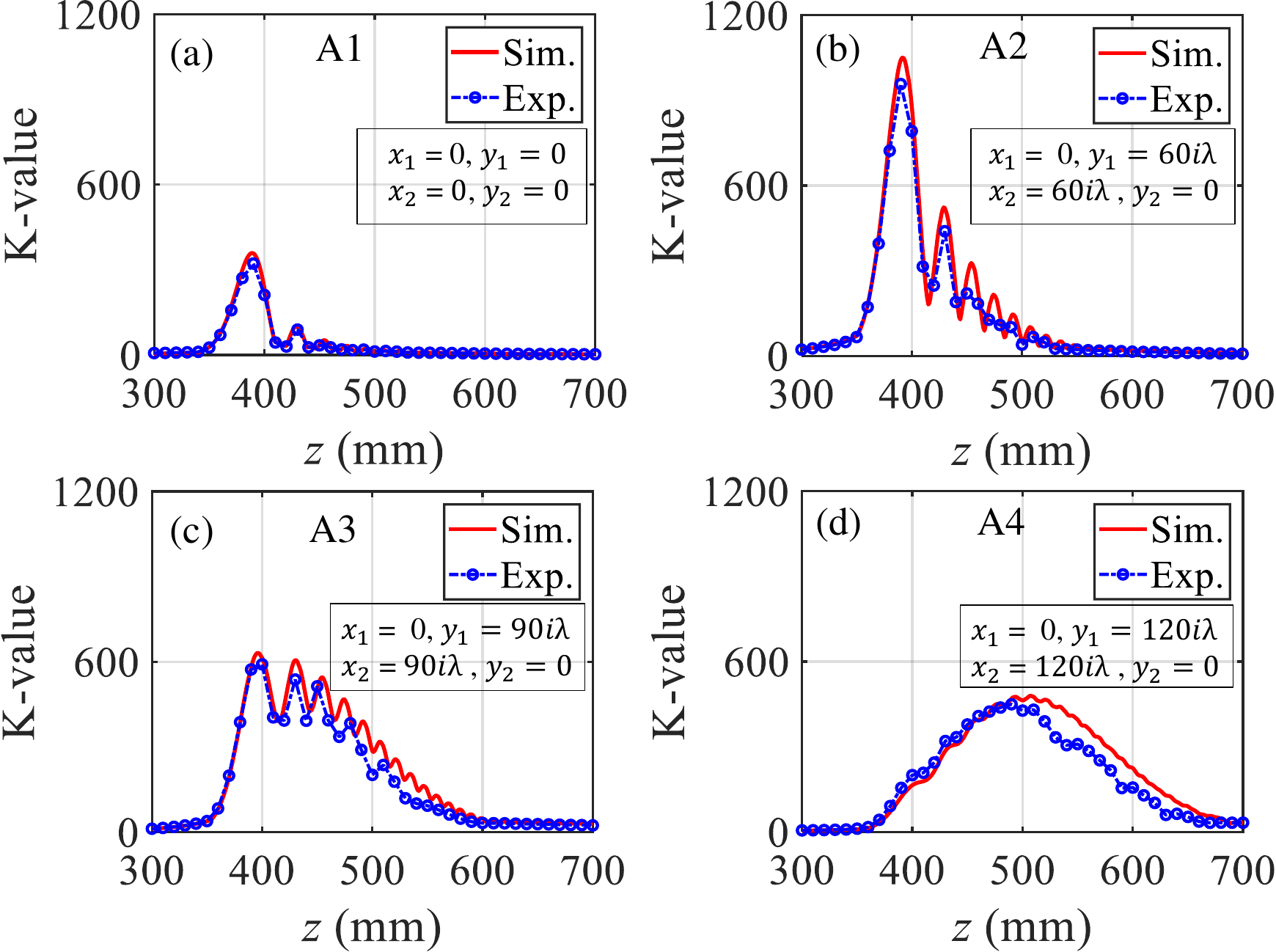}
\caption{Experimental (dashed blue curve with circles) and simulated (solid red curve) variation of the $\mathrm{K}$-value with propagation distance $z$ for ACADBs with the asymmetry cases (a) A1, (b) A2, (c) A3, and (d) A4.}
\label{fig:4}
\end{figure}

The autofocusing ability of ACADBs is quantified by the $\mathrm{K}$-value, defined as the ratio of the maximum intensity observed at the propagation plane ($\mathrm{I}_{\max}(z>0)$), to that at the initial plane ($\mathrm{I}_{\max}(z=0)$)~\cite{kumari2026abrupt,zang2022abruptly}. Figures\,\ref{fig:4}(a)–\ref{fig:4}(d) show the experimental and simulated variation of the $\mathrm{K}$-value as a function of $z$ correspond to A1 (symmetric), A2, A3, and A4 asymmetric cases, respectively.

For the symmetric case A1, the $\mathrm{K}$-value increases with $z$ as the beam intensity progressively concentrates toward the optical axis, reaching a maximum at $z_{\mathrm{af}} = 389\thinspace$mm with maximum $\mathrm{K}$-value $\sim 358$. Beyond this point, the $\mathrm{K}$-value decreases and subsequently exhibits secondary oscillations corresponding to secondary focal events. As the asymmetry increases, the intensity distribution at the initial plane is modified (Figs.\,\ref{fig:1}~(b)~–~\ref{fig:1}~(d)), and this leads to changes in the beam’s propagation. For A2, the maximum $\mathrm{K}$-value significantly increases to $\sim 1050$ at $z_{\mathrm{af}} = 392\thinspace$mm, indicating enhanced autofocusing ability. With further increase in asymmetry (A3 and A4), the maximum $\mathrm{K}$-values are found to be $\sim 629$ at $z_{\mathrm{af}} = 397\thinspace$mm and $\sim 473$ at $z_{\mathrm{af}} = 495\thinspace$mm, respectively. The shift in $z_{\mathrm{af}}$ suggests that increasing asymmetry alters the effective interference and phase redistribution mechanism governing the focusing dynamics. The analyzed maximum $\mathrm{K}$-values show that as the asymmetry increases, the autofocusing ability initially improves and then gradually decreases. The maximum enhancement is achieved for the A2 case, where the $\mathrm{K}$-value increases by approximately 193\thinspace\% compared to the symmetric case (A1). 

\begin{figure}[htbp]
\centering
\includegraphics[height = 3.2cm, keepaspectratio = true]{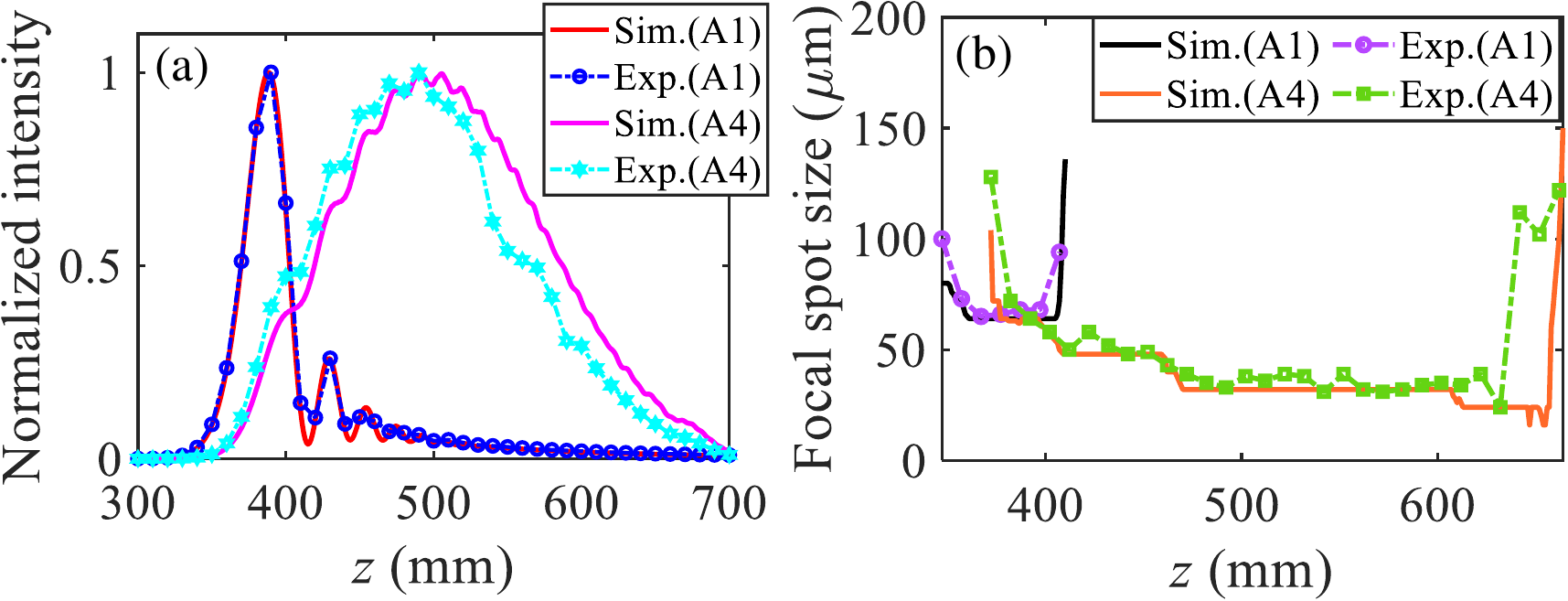}
\caption{(a) Normalized on-axis intensity variation of ACADBs with $z$: experimental (dashed blue curve with circles) and simulated (solid red curve) for A1 case, and experimental (dashed cyan curve with stars) and simulated (solid pink curve) for A4 case. (b) Focal spot size variation with $z$: experimental (dashed violet curve with circles) and simulated (solid black curve) for A1 case, and experimental (dashed green curve with squares) and simulated (solid orange curve) for A4 case.}
\label{fig:5}
\end{figure}

Further, to quantify the DOF, we have calculated the focal spot size and analyzed the variation of the on-axis normalized intensity as a function of $z$ ~\cite{kumari2024generating,chen2024optimizing}. The focal spot size of the beam is calculated by determining the full width at half maximum (FWHM) of the transverse intensity cross-sections of the central lobe at different propagation distances for each asymmetry case. Note that, for the asymmetric cases (A2--A4), the focal spot becomes elliptical due to the introduced complex transverse shifts. Therefore, the focal spot size is determined by averaging the FWHM of the intensity cross-sections measured along the two diagonals (see Appendix\,\ref{Appendix_B}). Furthermore, we define the focal region along the axial distance up to a cutoff where the on-axis normalized intensity remains $\geq 50\thinspace\%$. For this, the FWHM of the on-axis normalized intensity variation is considered, and the focal spot size is calculated over this axial distance. Figure\,\ref{fig:5}~(a) shows the experimental and simulated on-axis normalized intensity variation, and Fig.\,\ref{fig:5}(b) shows the corresponding calculated focal spot size as a function of $z$ for the A1 and A4 asymmetry cases (for A2 and A3 cases, see Appendix\,\ref{Appendix_B}).

From the variation of the normalized on-axis intensity (Fig.\,\ref{fig:5}~(a)) and K-value (Fig.\,\ref{fig:4}) as the asymmetry increases from A1 to A4, it is observed that the intensity associated with the secondary focal events gradually increases. In the A4 case, the peak intensity of the secondary focusing event becomes nearly comparable to that of the primary autofocusing event. This behavior indicates a significant extension of the axial length over which the on-axis normalised intensity is $\geq 50~\%$. Now, the focal spot size over the axial distance, within which it remains nearly constant and the normalized on-axis intensity stays $\geq 50\thinspace\%$, increases with asymmetry from A1 to A4. As evident, the DOF increases with asymmetry, where the focal spot size remains nearly constant over an extended axial range. The corresponding values are approximately $33\thinspace\mathrm{mm}$, $33\thinspace\mathrm{mm}$, $106\thinspace\mathrm{mm}$, and $166\thinspace\mathrm{mm}$ for A1, A2, A3, and A4 cases, respectively.

\begin{figure}[htbp]
\centering
\includegraphics[height = 3cm, keepaspectratio = true]{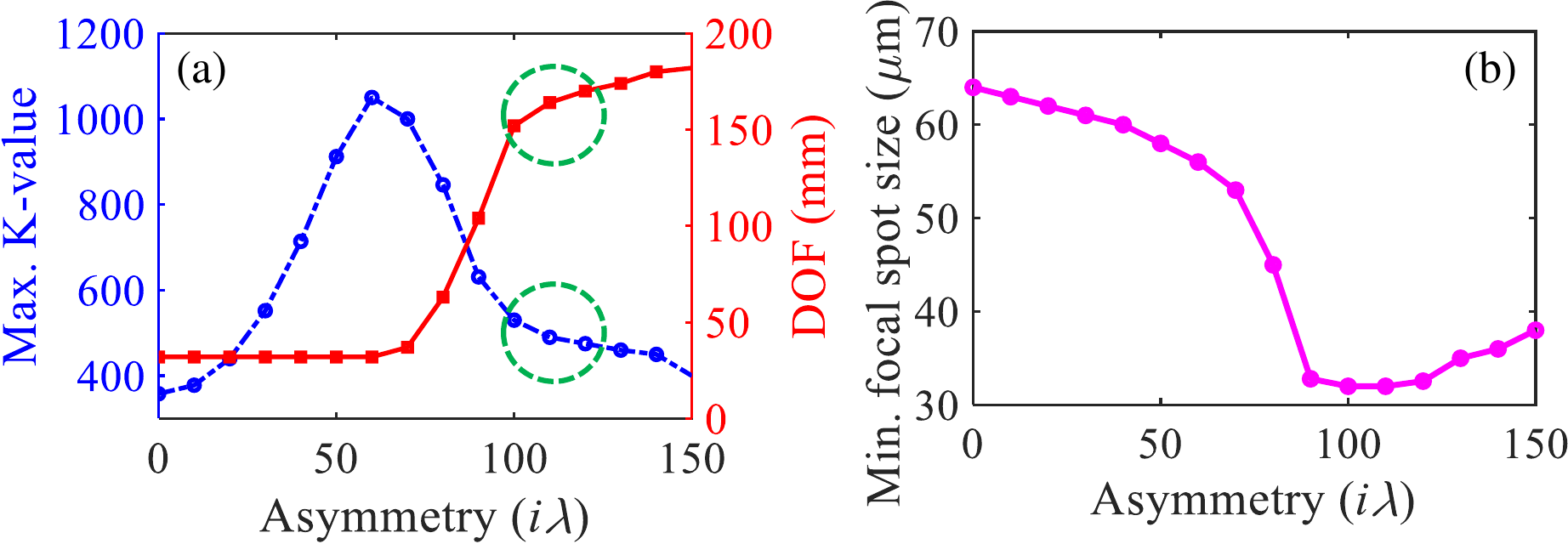}
\caption{(a) Variation of maximum $\mathrm{K}$-value (dashed blue curve with circles) and DOF (solid red curve with squares) as a function of asymmetry. (b) Variation of minimum focal spot size with asymmetry.}
\label{fig:6}
\end{figure}

Furthermore, the variation of the maximum $\mathrm{K}$-value, DOF, and minimum focal spot size as a function of asymmetry is shown in Fig.\,\ref{fig:6}, highlighting the trade-off and possible simultaneous optimization between autofocusing ability and DOF. As the asymmetry increases, the maximum $\mathrm{K}$-value initially increases, reaches an optimum, and subsequently decreases 
(Fig.\,\ref{fig:6}~(a)). In contrast, the DOF remains nearly constant in the initial regime, as the normalized on-axis intensity of the secondary focal events stays below $50~\%$ of the primary autofocusing peak and therefore does not extend the effective axial range of the focal region. However, beyond a certain level of asymmetry, the secondary focal peaks become sufficiently strong to contribute to the overall axial intensity envelope. As a result, the longitudinal intensity distribution becomes broader, leading to an extended axial region over which the beam maintains significant intensity and nearly constant focal spot size, and hence a corresponding increase in the DOF. As shown in Fig.\,\ref{fig:6}(b), the minimum focal spot size decreases from $64\thinspace\mu$m to $32\thinspace\mu$m with increasing asymmetry, with corresponding values of 64~$\mu$m, 56~$\mu$m, 33~$\mu$m, and 32~$\mu$m for the A1, A2, A3, and A4 cases, respectively. 

We also evaluate the focusing efficiency $\eta$, defined as the ratio of the optical power contained within a circular region of diameter equal to three times the focal spot size to the total incident power (see Appendix\,\ref{Appendix_C}) \cite{kumari2024generating}. The results show that, with increasing asymmetry, the peak value of $\eta$ gradually decreases, with corresponding values of $30\thinspace\%$, $26\thinspace\%$, $22.5\thinspace\%$, and $20\thinspace\%$ for the A1, A2, A3, and A4 cases, respectively; however, the axial length over which $\eta$ remains nearly constant increases significantly. This indicates that although the maximum focusing efficiency is reduced, the beam maintains a stable focusing performance over a longer propagation distance. Such behavior is consistent with the extension of DOF observed in the asymmetric configurations. Therefore, an optimal degree of asymmetry is required to achieve a simultaneous enhancement of autofocusing strength and DOF along with the smallest focal spot size while maintaining sufficiently high $\eta$. This makes the beam particularly suitable for potential applications. The optimal region, highlighted by the green dashed circles in Fig.\,\ref{fig:6}(a), corresponds approximately to complex transverse shifts in the range of $100\,i\lambda$~--~$120\,i\lambda$. This trade-off highlights the importance of controlled beam engineering for practical applications.

For comparison, we have also evaluated the Rayleigh range of a Gaussian beam with a size ($\mathrm{FWHM}\sim 32\,\mu\mathrm{m}$) corresponding to the minimum focal spot size of the A4 case ($32\thinspace\mu\mathrm{m}$), which is about $2.18\thinspace\mathrm{mm}$, whereas the ACADB exhibits of approximately $118\thinspace\mathrm{mm}$, i.e., about $\sim 54$ times larger than that of the corresponding Gaussian beam (see Appendix\,\ref{Appendix_D}).

\section{Conclusions}
\noindent
In summary, we have demonstrated a systematic enhancement of both abrupt autofocusing ability and DOF of CADBs by introducing controlled asymmetry in the initial plane. However, excessive asymmetry leads to a reduction in both maximum $\mathrm{K}$-value and $\eta$ due to intensity redistribution away from the central lobe. This reveals an inherent trade-off between autofocusing strength, DOF extension, and power concentration. Therefore, an optimal degree of asymmetry is essential to achieve simultaneous enhancement of axial confinement and focusing performance. A good agreement is found between the simulation and experimental results. These optimized ACADBs provide continuous focal regions over extended propagation distances, making them promising for applications in optical trapping, biomedical imaging, microfabrication, and optical communications.

\section*{Acknowledgements}
\noindent
We acknowledge financial support through the National Quantum Mission (NQM) of the Department of Science and Technology, Government of India. Anita Kumari acknowledges the fellowship support from the University Grants Commission (UGC).

\appendix
\section{Details of computer-generated hologram (CGH)} \label{Appendix_A}
\noindent
To generate the CGH, we numerically compute the Fourier transform of the electric field distribution of the ACADB (given by Eq.\,(\ref{Eq:1}) in the manuscript), which can be expressed as
\begin{equation}
\psi(\rho,\theta)=\mathcal{F}[E_{c}(r,\phi,0)]=A(\rho,\theta)\exp[i\zeta(\rho,\theta)],\label{eqA1}
\end{equation}
where $\mathcal{F}$ denotes the Fourier transform operation, $A(\rho,\theta)$ represents the normalized amplitude distribution, and $\zeta(\rho,\theta)$ represents the phase distribution.

Next, the complex field $\psi(\rho,\theta)$ is encoded into a phase-only CGH by using an appropriate phase transmittance function. This approach enables the incorporation of both amplitude and phase variations into a phase-only modulation format suitable for the SLM. The transmittance function is written as
\begin{equation}
  T(x,y)=\exp[i\Omega(A,\zeta)],\label{A2}
\end{equation}
where $\Omega(A,\zeta)$ accounts for both amplitude and phase information and corresponds to the complex field $\psi(\rho,\theta)$. To determine $\Omega(A,\zeta)$, we adopt a method reported in \cite{Arrizon:07}. The phase function $\Omega(A,\zeta)$ with odd symmetry can be expressed as
\begin{equation}
\Omega(A,\zeta)=f(A)\sin(\zeta), \label{A3}
\end{equation}
where $f(A)$ is an unknown function to be determined. Substituting this into the transmittance function and applying the Jacobi–Anger expansion, the phase transmittance can be expressed as a Fourier series
\begin{equation}
T(x,y)=\exp[i.f(A)\sin(\zeta)]=\sum_{l=-\infty}^{\infty} J_{l}[f(A)]\exp(il\zeta), \label{A4}
\end{equation}
where $J_{l}$ denotes the Bessel function of the first kind of order $l$. The desired complex field $\psi(\rho,\theta)$ is reconstructed from the first-order diffraction term ($l=1$) in series expansion, provided that the following condition is satisfied 
\begin{equation}
cA=J_{1}[f(A)],\label{A5}    
\end{equation}
where $c$ is a positive constant. The function $f(A)$ is obtained numerically by inverting this relation. From the series expansion, it is clear that the transmittance function generates multiple diffraction orders. However, only the first-order term is required for beam reconstruction. Therefore, the undesired diffraction orders ($l\neq1$) must be filtered out. To achieve this, a blazed grating is added to the phase pattern, which spatially separates the different diffraction orders. The modified phase pattern can be written as $\Omega(A,\zeta+2\pi(f_{x}x+f_{y}y))$, where $f_{x}$ and $f_{y}$ are the spatial carrier frequencies along the $x$ and $y$ directions, respectively.

As a result, the light reflected from the SLM contains several spatially separated diffraction orders. The desired first-order beam is then selected using a circular aperture (CA), ensuring accurate reconstruction of the required field.
\section{Analysis of focal spot size and the on-axis intensity distribution as functions of
propagation distance for the A2 and A3 asymmetry cases} \label{Appendix_B}
\noindent
We evaluate the focal spot size of the beam by measuring the full width at half maximum (FWHM) of the transverse intensity cross-sections of the central lobe at the autofocusing plane.
\begin{figure}[htbp]
\centering
\includegraphics[height = 4cm, keepaspectratio = true]{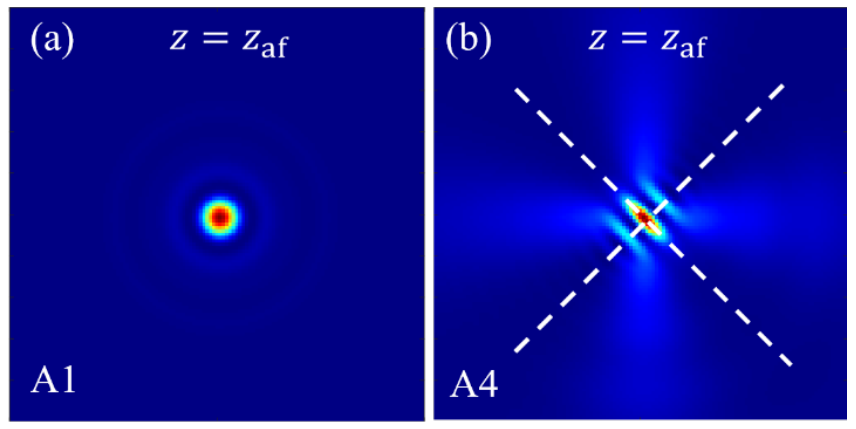}
\caption{Intensity distribution of the beam at the autofocusing plane for the cases of (a) A1 and (b) A4. The dashed white lines indicate the directions used to obtain the intensity cross-sections for calculating the FWHM in case of A4.}
\label{fig:7}
\end{figure}
Figure\,\ref{fig:7} shows the intensity distributions at the autofocusing point for cases A1 (Fig.\,\ref{fig:7}(a)) and A4 (Fig.\,\ref{fig:7}(b)). For the symmetric case (A1), the focal spot has a circular shape, therefore a single transverse intensity cross-section is sufficient to determine the FWHM. However, for the asymmetric case the focal spot becomes elliptical. In this situation, two intensity cross-sections are taken along the diagonals indicated by the dashed white lines, and the focal spot size is obtained by averaging the corresponding FWHM values.

\begin{figure}[htbp]
\centering
\includegraphics[height = 3.34cm, keepaspectratio = true]{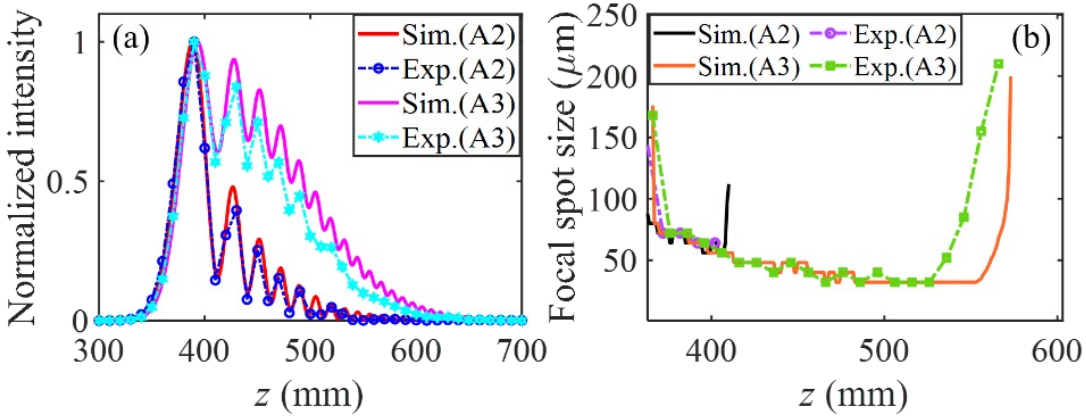}
\caption{(a) Normalized on-axis intensity variation of ACADBs with propagation distance z: experimental (dashed dark blue curve with circles) and simulated (solid red curve) for A2 case, and experimental (dashed blue curve with stars) and simulated (solid pink curve) for A3 case. (b) Variation of focal spot size with propagation distance z: experimental (dashed violet curve with circles) and simulated (solid black curve) for A2 case, and experimental (dashed green curve with squares) and simulated (solid orange curve) for A3 case.}
\label{fig:8}
\end{figure}

In Fig.\,\ref{fig:8}(a) and \ref{fig:8}(b), we present the experimental and simulated results, where Fig.\,\ref{fig:8}(a) shows the variation of the on-axis normalized intensity distribution with propagation distance and Fig.\,\ref{fig:8}(b) shows the corresponding variation of the focal spot size with propagation distance for the A2 and A3 asymmetry cases.

\vspace{-20pt}
\section{Focusing efficiency with propagation distance} \label{Appendix_C}
\begin{figure}[htbp]
\centering
\includegraphics[height = 5cm, keepaspectratio = true]{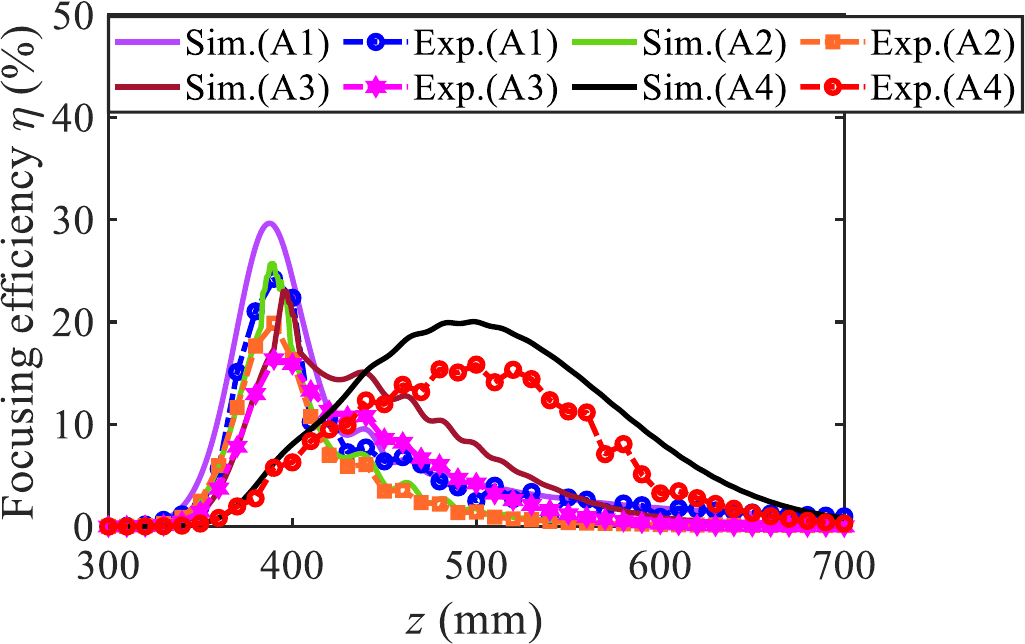}
\caption{Experimental (dashed blue curve with circles) and simulated (solid violet curve) variation of focusing efficiency $\eta$ (\%) with propagation distance z for A1, experimental (dashed orange curve with squares) and simulated (solid green curve) results for A2, experimental (dashed pink curve with stars) and simulated (solid brown curve) results for A3, and experimental (dashed red curve with circles) and simulated (solid black curve) results for A4.}
\label{fig:9}
\end{figure}

\noindent
In Fig.\,\ref{fig:9}, we present the measured variation of the focusing efficiency $\eta$ (\%) with propagation distance for all the asymmetry cases considered in the manuscript. 
The results show that as the introduced asymmetry increases, the peak value of the focusing efficiency gradually decreases. However, the propagation distance over which the efficiency remains nearly constant increases as we move from case A1 to A4. This indicates that although the maximum intensity concentration at the primary focus becomes slightly weaker with increasing asymmetry, the beam maintains a relatively stable intensity distribution over a longer axial region. Such behavior is consistent with the increase in the DOF, since the intensity remains distributed over an extended propagation distance rather than being strongly localized at a single focal point. Consequently, the introduction of controlled asymmetry reduces the peak efficiency but contributes to an axial extension of the effective focusing region, which supports the enhanced DOF observed in the asymmetric cases.
\vspace{-10pt}
\section{Depth-of-focus comparison between Gaussian and ACADBs} \label{Appendix_D}
\noindent
For comparison of the DOF between a Gaussian beam and ACADBs, we consider a Gaussian beam
whose beam waist $w$ corresponds to the minimum focal spot size obtained for the A4 asymmetry case. 
\begin{figure}[htbp]
\centering
\includegraphics[height = 7cm, keepaspectratio = true]{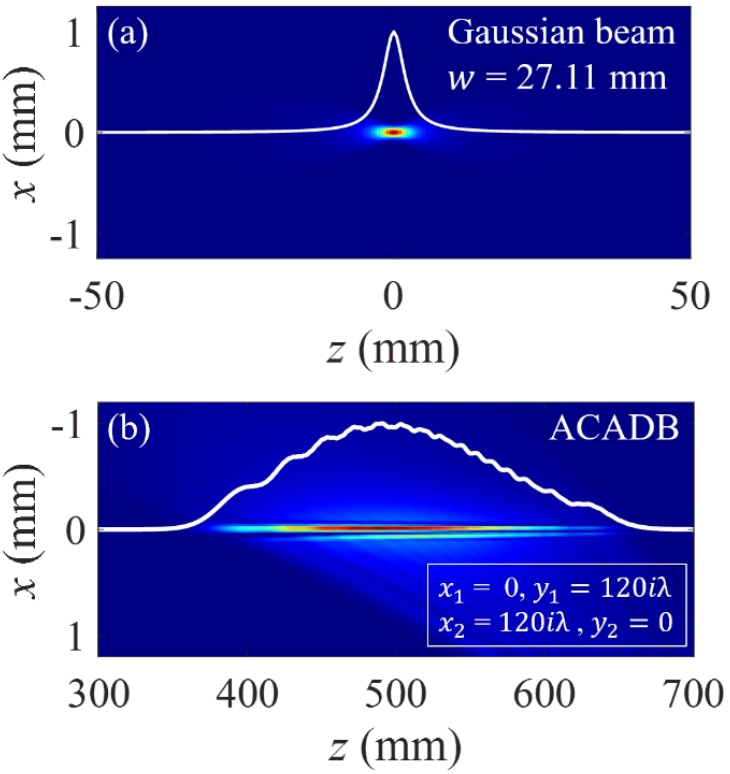}
\caption{Simulated intensity distribution in the x-z plane corresponding to (a) a Gaussian beam and, (b) ACADBs (A4 case). Note, here the white curve represents the on-axis intensity variation.}
\label{fig:10}
\end{figure}
The minimum focal spot size in this case is approximately $32\thinspace\mu$m, which corresponds to a Gaussian beam waist of about $27.11\thinspace\mu$m (FWHM = $32\thinspace\mu$m). Using this waist, the Gaussian beam is propagated and its Rayleigh range is determined by the axial distance over which the beam waist increases to $\sqrt{2}\omega$. From this criterion, the Rayleigh range of the Gaussian beam is found to be approximately $2.16\thinspace$mm (Fig.\,\ref{fig:10}(a)). In contrast, the ACADB exhibits a much longer axial length of about $118\thinspace$mm (Fig.\,\ref{fig:10}(b)), which is also calculated similarly as the Rayleigh range in a Gaussian beam propagation. This indicates that the ACADB provides nearly $54$ times larger than the Rayleigh range of the corresponding Gaussian beam, demonstrating its significantly enhanced DOF.
\bibliography{sample}{}

@article{zhang2011trapping,
  title={Trapping and guiding microparticles with morphing autofocusing Airy beams},
  author={Zhang, Peng and Prakash, Jai and Zhang, Ze and Mills, Matthew S and Efremidis, Nikolaos K and Christodoulides, Demetrios N and Chen, Zhigang},
  journal={Optics Letters},
  volume={36},
  number={15},
  pages={2883--2885},
  year={2011},
  publisher={Optical Society of America}
}

@article{kafian2020light,
  title={Light-sheet fluorescence microscopy with scanning non-diffracting beams},
  author={Kafian, Hosein and Lalenejad, Meelad and Moradi-Mehr, Sahar and Birgani, Shiva Akbari and Abdollahpour, Daryoush},
  journal={Sci. Rep.},
  volume={10},
  number={1},
  pages={8501},
  year={2020},
  publisher={Nature Publishing Group UK London}
}

@article{lorenser2012ultrathin,
  title={Ultrathin fiber probes with extended depth of focus for optical coherence tomography},
  author={Lorenser, Dirk and Yang, Xiaojie and Sampson, David D},
  journal={Optics Letters},
  volume={37},
  number={10},
  pages={1616--1618},
  year={2012},
  publisher={Optical Society of America}
}

@article{planchon2011rapid,
  title={Rapid three-dimensional isotropic imaging of living cells using Bessel beam plane illumination},
  author={Planchon, Thomas A and Gao, Liang and Milkie, Daniel E and Davidson, Michael W and Galbraith, James A and Galbraith, Catherine G and Betzig, Eric},
  journal={Nature Methods},
  volume={8},
  number={5},
  pages={417--423},
  year={2011},
  publisher={Nature Publishing Group US New York}
}

@article{fu2024extending,
  title={Extending femtosecond laser superfilamentation in air with a multifocal phase mask},
  author={Fu, Silin and Mysyrowicz, Andr{\'e} and Arantchouk, Leonid and Lozano, Magali and Houard, Aur{\'e}lien},
  journal={Applied Physics Letters},
  volume={125},
  number={1},
  year={2024},
  publisher={AIP Publishing}
}

@article{killinger2002free,
  title={Free space optics for laser communication through the air},
  author={Killinger, Dennis},
  journal={OPN},
  volume={13},
  number={10},
  pages={36--42},
  year={2002},
  publisher={OSA}
}

@article{khonina2020bessel,
  title={Bessel beam: significance and applications—a progressive review},
  author={Khonina, Svetlana Nikolaevna and Kazanskiy, Nikolay Lvovich and Karpeev, Sergey Vladimirovich and Butt, Muhammad Ali},
  journal={Micromachines},
  volume={11},
  number={11},
  pages={997},
  year={2020},
  publisher={MDPI}
}

@article{efremidis2019airy,
  title={Airy beams and accelerating waves: an overview of recent advances},
  author={Efremidis, Nikolaos K and Chen, Zhigang and Segev, Mordechai and Christodoulides, Demetrios N},
  journal={Optica},
  volume={6},
  number={5},
  pages={686--701},
  year={2019},
  publisher={Optical Society of America}
}

@article{zhao2022flexible,
  title={Flexible method for generating needle-shaped beams and its application in optical coherence tomography},
  author={Zhao, Jingjing and Winetraub, Yonatan and Du, Lin and Van Vleck, Aidan and Ichimura, Kenzo and Huang, Cheng and Aasi, Sumaira Z and Sarin, Kavita Y and de la Zerda, Adam},
  journal={Optica},
  volume={9},
  number={8},
  pages={859--867},
  year={2022},
  publisher={Optica Publishing Group}
}

@article{hu2022experimental,
  title={Experimental demonstration of a “pin-like” low-divergence beam in a 1-Gbit/s OOK FSO link using a limited-size receiver aperture at various propagation distances},
  author={Nanzhe Hu and Huibin Zhou and Runzhou Zhang and Haoqian Song and Kai Pang and Kaiheng Zou and Hao Song and Xinzhou Su and Cong Liu and Brittany Lynn and Moshe Tur and Alan E. Willner},
  journal={Optics Letters},
  volume={47},
  number={16},
  pages={4215--4218},
  year={2022},
  publisher={Optica Publishing Group}
}

@article{lu2019abruptly,
  title={Abruptly autofocusing property and optical manipulation of circular Airy beams},
  author={Lu, Wanli and Sun, Xu and Chen, Huajin and Liu, Shiyang and Lin, Zhifang},
  journal={Physical Review A},
  volume={99},
  number={1},
  pages={013817},
  year={2019},
  publisher={APS}
}

@article{jiang2015propagation,
  title={Propagation characteristics of the modified circular Airy beam},
  author={Jiang, Yunfeng and Zhu, Xiuwei and Yu, Wenlei and Shao, Hehong and Zheng, Wanting and Lu, Xuanhui},
  journal={Optics Express},
  volume={23},
  number={23},
  pages={29834--29841},
  year={2015},
  publisher={Optical Society of America}
}

@article{zang2022abruptly,
  title={Abruptly autofocusing of generalized circular Airy derivative beams},
  author={Zang, Xiang and Dan, Wensong and Zhou, Yimin and Lv, Han and Wang, Fei and Cai, Yangjian and Zhou, Guoquan},
  journal={Optics Express},
  volume={30},
  number={3},
  pages={3804--3819},
  year={2022},
  publisher={Optica Publishing Group}
}

@article{he2023key,
  title={Key to an extension or shortening of focal length in the enhancement of autofocusing ability of a circular Airyprime beam caused by a linear chirp factor},
  author={He, Jian and Zang, Xiang and Dan, Wensong and Zhou, Yimin and Wang, Fei and Cai, Yangjian and Zhou, Guoquan},
  journal={Optics Letters},
  volume={48},
  number={9},
  pages={2365--2368},
  year={2023},
  publisher={Optica Publishing Group}
}

@article{zang2023simultaneously,
  title={Simultaneously enhancing autofocusing ability and extending focal length for a ring Airyprime beam array by a linear chirp},
  author={Zang, Xiang and Dan, Wensong and Zhou, Yimin and Wang, Fei and Cai, Yangjian and Zhou, Guoquan},
  journal={Optics Letters},
  volume={48},
  number={4},
  pages={912--915},
  year={2023},
  publisher={Optica Publishing Group}
}

@article{he2024propagation,
  title={Propagation characteristics of a ring Airyprime vortex beam and an Airyprime vortex beam array in atmospheric turbulence},
  author={He, Jian and Dan, Wen-Song and Chen, Jia-Hao and Wang, Fei and Zhou, Yi-Min and Zhou, Guo-Quan},
  journal={Result. Phys.},
  volume={62},
  pages={107827},
  year={2024},
  publisher={Elsevier}
}

@article{kumari2024autofocusing,
  title={Autofocusing and self-healing of partially blocked circular Airy derivative beams},
  author={Kumari, Anita and Dev, Vasu and Pal, Vishwa},
  journal={Opt. Laser Technol.},
  volume={168},
  pages={109837},
  year={2024},
  publisher={Elsevier}
}

@article{kumari2025abrupt,
  title={Abrupt autofocusing of circular Airy derivative beams in complex media},
  author={Kumari, Anita and Dev, Vasu and Pal, Vishwa},
  journal={Opt. Laser Technol.},
  volume={183},
  pages={112319},
  year={2025},
  publisher={Elsevier}
}

@article{kumari2026abrupt,
  title={Abrupt autofocusing and scintillation dynamics of truncated circular airy derivative beams in strong turbulence},
  author={Kumari, Anita and Pal, Vishwa},
  journal={Optics Communications},
  volume={607},
  pages={132923},
  year={2026},
  publisher={Elsevier}
}

@article{zheng2024adjustable,
  title={Adjustable focusing property of circular Airyprime beam through Fourier space modulation},
  author={Zheng, Xinqing and Yang, Yongzheng and Liu, Yejin and Lin, Xiaojun and Liang, Zehong and Liu, Jie and Deng, Dongmei},
  journal={Optics Letters},
  volume={49},
  number={15},
  pages={4393--4396},
  year={2024},
  publisher={Optica Publishing Group}
}

@article{collins1970,
  title={Lens-system diffraction integral written in terms of matrix optics},
  author={Collins, Stuart A},
  journal={Journal of Optical Society of America},
  volume={60},
  number={9},
  pages={1168--1177},
  year={1970},
  publisher={Optica Publishing Group}
}

@article{kumari2024generating,
  title={Generating optical vortex needle beams with a flat diffractive lens},
  author={Kumari, Anita and Dev, Vasu and Hayward, Tina M and Menon, Rajesh and Pal, Vishwa},
  journal={Journal of Applied Physics},
  volume={136},
  number={11},
  year={2024},
  publisher={AIP Publishing}
}

@article{chen2024optimizing,
  title={Optimizing airy needle-like beams for long-range axial manipulation and super-resolution imaging},
  author={Chen, Lai and Gao, Binjie and Li, Xiao and Chan, CT and Wang, Jun and Ye, Linhua and Wang, Li-Gang},
  journal={ACS Photonics},
  volume={11},
  number={9},
  pages={3610--3620},
  year={2024},
  publisher={ACS Publications}
}

@article{Papazoglou:11,
author = {Dimitrios G. Papazoglou and Nikolaos K. Efremidis and Demetrios N. Christodoulides and Stelios Tzortzakis},
journal = {Opt. Lett.},
number = {10},
pages = {1842--1844},
publisher = {Optica Publishing Group},
title = {Observation of abruptly autofocusing waves},
volume = {36},
month = {May},
year = {2011},
}

@article{Chen:18,
author = {Xingyu Chen and Dongmei Deng and Jingli Zhuang and Xi Peng and Dongdong Li and Liping Zhang and Fang Zhao and Xiangbo Yang and Hongzhan Liu and Guanghui Wang},
journal = {Opt. Lett.},
number = {15},
pages = {3626--3629},
publisher = {Optica Publishing Group},
title = {Focusing properties of circle Pearcey beams},
volume = {43},
month = {Aug},
year = {2018},
}

@article{Ring:12,
author = {James D. Ring and Jari Lindberg and Areti Mourka and Michael Mazilu and Kishan Dholakia and Mark R. Dennis},
journal = {Opt. Express},
number = {17},
pages = {18955--18966},
publisher = {Optica Publishing Group},
title = {Auto-focusing and self-healing of Pearcey beams},
volume = {20},
month = {Aug},
year = {2012},
}

@article{Chen2:18,
author = {Xingyu Chen and Dongmei Deng and Jingli Zhuang and Xiangbo Yang and Hongzhan Liu and Guanghui Wang},
journal = {Appl. Opt.},
number = {28},
pages = {8418--8423},
publisher = {Optica Publishing Group},
title = {Nonparaxial propagation of abruptly autofocusing circular Pearcey Gaussian beams},
volume = {57},
month = {Oct},
year = {2018},
}

@article{Zang:22,
author = {Xiang Zang and Wensong Dan and Fei Wang and Yimin Zhou and Yangjian Cai and Guoquan Zhou},
journal = {Opt. Lett.},
number = {21},
pages = {5654--5657},
publisher = {Optica Publishing Group},
title = {Dependence of autofocusing ability of a ring Airyprime beams array on the number of beamlets},
volume = {47},
month = {Nov},
year = {2022},
}

@article{Arrizon:07,
author = {Victor Arriz\'{o}n and Ulises Ruiz and Rosibel Carrada and Luis A. Gonz\'{a}lez},
journal = {J. Opt. Soc. Am. A},
number = {11},
pages = {3500--3507},
publisher = {Optica Publishing Group},
title = {Pixelated phase computer holograms for the accurate encoding of scalar complex fields},
volume = {24},
month = {Nov},
year = {2007},
}
\bibliographystyle{apsrev4-1}
\end{document}